\documentclass[twocolumn,      
        showpacs,              
        preprintnumbers,       
        aps,                   
        prd,                   
        a4paper,               
        superscriptaddress,    
        unsortedaddress,       
        footinbib,             
        tightenlines,          
        floats                 
        ]{revtex4} 
 
\usepackage{graphicx} 
\usepackage{latexsym} 
\usepackage{amsmath,amssymb}   
\usepackage[draft=false]{hyperref}

\newcommand{\de}{\mathrm{d}} 
\renewcommand{\(}{\left(} 
\renewcommand{\)}{\right)}

\newcommand{\period}{\,\mathrm{.}} 
\newcommand{\comma}{\,\mathrm{,}} 
\newcommand{\reffig}[1]{Fig.~\ref{#1}} 
\newcommand{\refeq}[1]{Eq.~(\ref{#1})} 
\newcommand{\reftab}[1]{Table~\ref{#1}} 
 
\newcommand{\lag}{\mathcal{L}} 
\newcommand{\abs}[1]{\left\vert#1\right\vert} 
 
\newcommand{\Order}{\mathcal{O}} 
 
\newcommand{\pti}{\tilde{p}} 
\newcommand{\eti}{\tilde{\varepsilon}} 
\newcommand{\ek}{\varepsilon_\mathrm{k}} 
\newcommand{\ef}{\varepsilon_\mathrm{f}} 
\newcommand{\pk}{p_\mathrm{k}} 
\newcommand{\wk}{w_\mathrm{k}} 
\newcommand{\wf}{w_\mathrm{f}} 
\newcommand{\csk}{c^2_\mathrm{sk}} 
\newcommand{\Ok}{\Omega_\mathrm{k}} 
\newcommand{\Okc}{\Omega_\mathrm{kc}}
\newcommand{\Oks}{\Omega_\mathrm{ks}}
\newcommand{\yc}{y_\mathrm{c}}
\newcommand{\ys}{y_\mathrm{s}} 
\newcommand{\dy}{\delta y} 
\newcommand{\Xc}{X_c} 
\newcommand{\Xw}{X_w} 
\newcommand{\Xs}{X_\star} 
\newcommand{\R}{\mathbf{R}} 
\newcommand{\K}{\mathbf{K}} 
\newcommand{\xx}{\mathbf{x}} 
\newcommand{\OO}{\mathbf{0}} 
\newcommand{\So}{\mathbf{S_0}} 
\newcommand{\Sp}{\mathbf{S_+}} 
\newcommand{\Sm}{\mathbf{S_-}} 
 
\begin{document}


\title{K-essence and the coincidence problem} 
\author{Micha\"el Malquarti} 
\affiliation{Astronomy Centre, University of Sussex, 
       Brighton BN1 9QJ, United Kingdom} 
\author{Edmund J. Copeland} 
\affiliation{Centre for Theoretical Physics, University of Sussex, 
       Brighton BN1 9QJ, United Kingdom} 
\author{Andrew R. Liddle} 
\affiliation{Astronomy Centre, University of Sussex, 
       Brighton BN1 9QJ, United Kingdom} 
\date{\today} 
\pacs{98.80.Cq \hfill astro-ph/0304277} 
\preprint{astro-ph/0304277}


\begin{abstract} 
K-essence has been proposed as a possible means of explaining the coincidence 
problem of the Universe beginning to accelerate only at the present epoch. We 
carry out a comprehensive dynamical systems analysis of the k-essence models 
given so far in the literature. We numerically study the basin of attraction of 
the tracker solutions and we highlight the behaviour of the field close to sound 
speed divergences. We find that, when written in terms of parameters with a 
simple dynamical interpretation, the basins of attraction represent only a small 
region of the phase space. 
\end{abstract} 
 
\maketitle


\section{Introduction}

The search for an explanation for the observed acceleration of the 
Universe~\cite{acce} is one of the most important challenges in contemporary 
cosmology. One usually assumes the existence of a dark energy component which 
breaks the strong energy condition, and the simplest model that one can build is 
that of introducing a cosmological constant. Yet, in order to obtain its 
domination today it has to be precisely set to an extremely small and so far 
unexplained value, one manifestation of the cosmological constant 
problem~\cite{W89}. More generally, the fact that the dark energy overtakes dark 
matter at a recent epoch ($z<1$) when we cosmologists are able to observe it is 
known as the coincidence problem. 

Allowing the dark energy to be dynamical may help to solve this issue. In this 
respect, a light scalar field, known as quintessence~\cite{quin,trac}, has been 
proposed. The field is meant to slow-roll down its potential, with its potential 
energy acting analogously to that of early Universe inflation. A wide class of 
tracker models~\cite{trac} feature an attractor solution which roughly mimics 
the behaviour of the dominant component of the Universe, rendering the evolution 
of the field fairly independent of its initial conditions. Unfortunately, in 
order to obtain quintessence domination today, the parameters of the potentials 
so far discussed also need a fine-tuning, and so as yet those models have not 
led to a compelling resolution of the coincidence problem. 

More recently, models based on scalar fields with non-canonical kinetic 
energy~\cite{kinf}, dubbed as k-essence~\cite{COY,AMS1,AMS2}, have emerged. A 
subclass of models~\cite{AMS1,AMS2} feature a tracker behaviour during radiation 
domination, and a cosmological-constant-like behaviour shortly after the 
transition to matter domination. As long as this transition seems to occur 
generically for purely dynamical reasons, these models are claimed to solve the 
coincidence problem without fine-tuning. 

In this paper we will analyze the models given in the literature so 
far~\cite{AMS1,AMS2}. We will study the size of the basin of attraction of 
their tracker solutions and comment on the fine-tuning of the parameters. We 
will also look at the behaviour of a general k-essence field close to 
singularities corresponding to a diverging sound speed.

Throughout this article a prime denotes a derivative with respect to the 
argument of the function to which it is applied, and a dot denotes a derivative 
with respect to proper time. We assume $3/8\pi G=1$.

\section{K-essence formalism} 
\label{formalism}

In general k-essence is defined as a scalar field with non-canonical kinetic 
energy, but usually the models are restricted to the Lagrangian 
\begin{equation} 
\lag_\mathrm{k} \equiv K(\phi)\pti(X) \comma 
\end{equation} 
where $K(\phi)>0$ and $X=\frac12\nabla_\mu\phi\nabla^\mu\phi$. We note that the 
definition includes quintessence models (in this paper meaning scalar fields 
with 
canonical kinetic term). Using the perfect fluid analogy, the pressure and the 
energy density are given by 
\begin{eqnarray} 
\pk(\phi,X) &=& K(\phi)\pti(X) \comma \\ 
\ek(\phi,X) &=& K(\phi)\eti(X) \comma 
\end{eqnarray} 
where 
\begin{equation} 
\eti(X) = 2X\pti'(X)-\pti(X) \period 
\end{equation} 
 
Following Refs.~\cite{AMS1,AMS2}, we set $K(\phi)\equiv1/\phi^2$, define a new 
variable $y\equiv1/\sqrt{X}$ and re-express $\pti(X)$ as $\pti(X(y))\equiv 
g(y)/y$. In this case the pressure and the energy density become 
\begin{eqnarray} 
\pk(\phi,y) &=& \frac{g(y)}{\phi^2y} \comma \\ 
\ek(\phi,y) &=& -\frac{g'(y)}{\phi^2} \period 
\end{eqnarray} 
We assume $\ek(\phi,y)>0$, hence $g'(y)<0$. The equation of state parameter and 
the effective sound speed are given by 
\begin{eqnarray} 
\label{wk} 
\wk(y) &=& -\frac{g(y)}{yg'(y)} \comma \\ 
\csk(y) &=& \frac{g(y)-yg'(y)}{y^2g''(y)} \period 
\label{csk} 
\end{eqnarray} 
We also assume $\wk(y)>-1$ and $\csk(y)>0$ which implies $g''(y)>0$. As a 
result, $g(y)$ must be a convex and decreasing function of $y$. 
 
{}From now on, we consider a flat Robertson--Walker Universe defined by the 
metric 
\begin{equation} 
\de s^2 = -\de t^2+a^2(t)\de\mathbf{x}^2 \period 
\end{equation} 
In this case, the Euler--Lagrange equation for the k-essence field is 
\begin{equation} 
\eti'(X)\ddot{\phi}+3H\pti'(X)\dot{\phi}+\frac{K'(\phi)}{K(\phi)}\eti(X)=0 
\comma 
\end{equation} 
where $H=\dot{a}/a$. Then, if the Universe is filled with another 
fluid with energy density $\ef$ and equation of state parameter 
$\wf$ constant one can find the following system of equations in 
terms of the independent variables $y$ and $\Ok$ 
\begin{eqnarray} 
\label{dybydn} 
\frac{\de y}{\de N} &=& 
\frac{\sqrt{-8g'(y)}\(r(y)-\sqrt{\Ok}\)}{yg''(y)} \comma \\ 
\frac{\de \Ok}{\de N} &=& 3\Ok\(1-\Ok\)\(\wf-\wk(y)\) \comma 
\end{eqnarray} 
where $N\equiv\ln(a/a_0)$, $\Ok\equiv\ek/(\ek+\ef)$ and 
\begin{equation} 
r(y) \equiv \frac{3\(g(y)-yg'(y)\)}{\sqrt{-8g'(y)}}>0 \period 
\end{equation} 
Here, we have assumed that $\dot{\phi}>0$. Therefore, for $0<y$ and $0<\Ok<1$, 
the dynamics is completely described by trajectories in the $y$--$\Ok$ plane. As 
long as $g''(y)\neq0$ the system is well defined. As we can see, $(\ys,\Oks)$ is 
a stationary point if $\wk(\ys)=\wf$ and $r^2(\ys)=\Oks$. As shown in 
Ref.~\cite{AMS2} this is a stable point --- and therefore corresponds to a 
perfect tracking ($\wk=\wf$) of the dominant fluid --- if $\csk(\ys)>\wk(\ys)$.

\section{Sound speed divergence}

As we will see, for the particular class of models we will analyze, we have 
$g''(\yc)=0$ for some $\yc$, which implies that $\eti'(\yc)=0$ and that the 
sound speed diverges at $\yc$. In that case, from Eq.~(\ref{dybydn}) we see that 
at $y=\yc$ there is a unique possible value $\Ok=\Okc$ given by the constraint 
equation
\begin{equation} 
\Okc=r^2(\yc) \period 
\end{equation} 
This means that the phase space $y$--$\Ok$ is cut into (at least) 
two parts separated by the line $y=\yc$ which is not allowed by 
the model except at $\So\equiv(\yc,\Okc)$. To study the dynamics close to this 
line we use the expansion $y=\yc+\dy$ with $\dy/\yc\ll1$, compute \refeq{dybydn} 
to $\Order(\dy^2)$ and find the equation
\begin{equation}
\label{perturbation}
\frac{\de\dy}{\de N} \simeq \frac{C_1}{\dy}+C_2\dy \comma
\end{equation}
where
\begin{eqnarray}
C_1 &=&
\frac{\sqrt{-8g'(\yc)}\(r(\yc)-\sqrt{\Ok}\)}{\yc g'''(\yc)}\\
C_2 &=&
\frac{3\(\wk(\yc)-1\)}{4}-\frac{2\(r(\yc)-\sqrt{\Ok}\)}{\yc\sqrt{-8g'(\yc)}}
\period
\end{eqnarray}
If $g'''(\yc)\neq0$, \refeq{perturbation} allows us to study the behaviour of 
the field close to the line $y=\yc$, and unless $\Ok=r^2(\yc)$ the first term 
dominates. For the particular class of models we will analyze we have 
$g'''(\yc)>0$, and therefore if $\Ok<r^2(\yc)$, hence $C_1>0$, the solution 
moves away from the line, whereas if $\Ok>r^2(\yc)$ the solution ceases to exist 
(in that it hits the singularity $\yc$) within a finite time $\Delta 
N\simeq\dy^2/2\abs{C_1}$ as it approaches the line (this has been checked 
numerically). We define $\Sp\equiv\{(\yc,\Ok)\vert\Ok<r^2(\yc)\}$ as the segment 
of the singularity from which some trajectories spontaneously emerge and 
$\Sm\equiv\{(\yc,\Ok)\vert\Ok>r^2(\yc)\}$ as the segment of the singularity on 
which some trajectories abruptly end. We can also use Eq.~(\ref{perturbation}) 
to determine the nature of the perturbation $\dy$ at the regular point 
$\So=(\yc,\Okc)$. In that case the equation simplifies to give 
\begin{equation} 
\label{perturbation1} 
\frac{\de\dy}{\de N} \simeq \frac{3}{4}(\wk(\yc)-1)\dy \period 
\end{equation} 
Depending on the value of $w(\yc)$ this solution either grows or decreases 
exponentially fast. In the cases we will study here, we have $\wk(\yc)<1$ and 
therefore $\dy$ decreases.

The existence of this singular behaviour means that a diverging sound speed 
leads to serious problems. In some situations it may well be possible to argue 
that the theory is valid up to a certain cut-off which excludes the 
singularity, but as we will show this is not the case for the 
cosmologically-realistic models proposed so far~\cite{AMS1,AMS2}. Instead, in 
these situations we must deal with the singular regions as we meet them.

\section{Analysis of two models} 
\label{analysis}

K-essence models can possess many different attractor 
solutions~\cite{COY,AMS1,AMS2}, especially trackers which perfectly mimic the 
dominant component of the Universe and attractors with negative equation of 
state leading to domination of the field. By choosing an appropriate function 
$g(y)$, it is possible to build a model with a certain number of attractor 
solutions which can feature some interesting dynamical properties. For instance, 
a stable tracker solution during radiation domination ($\R$) could render the 
late-time evolution of the field fairly independent of its initial conditions. 
Then, the lack of such a solution during matter domination would force the field 
to reach another pseudo-attractor with $\wk\approx-1$ leading to the k-essence 
domination attractor ($\K$). This possibility was discussed in 
Refs.~\cite{AMS1,AMS2} and in that sense it is possible to find a model which 
could solve the coincidence problem: we cosmologists would observe the 
acceleration of the Universe today because we happen to appear soon after the 
onset of matter domination which leads to the formation of structures --- and 
human beings. In this section we analyze the models proposed so far which are 
built in order to fulfill this principle~\cite{AMS1,AMS2}. We study their 
sensitivity to initial conditions and comment on the fine-tuning involved. 
Throughout the discussion we will refer to and make use of the parameters first 
introduced in Refs.~\cite{AMS1,AMS2}.

\subsection{Model 1 (Ref.~[7])}

Following the classification scheme introduced in Ref.~\cite{AMS2}, this first 
model is of type $(\mathrm{B}_r)$. It is defined by 
\begin{equation}
\label{model1}
\pti(X) \equiv -2.01+2\sqrt{1+X}+3\times10^{-17}X^3-10^{-24}X^4 \period 
\end{equation} 
As we will see, it appears to contain a number of problematic issues which on 
the face of it contradict what is stated in Ref.~\cite{AMS1}. For example for 
some ranges of $X$ we see that it is possible to have $\ek<0$, $\wk<-1$ and 
$\csk<0$. Moreover we find that $\csk$ and $\wk$ diverge at 
$X=\Xc\approx1.6\times10^7$ and $X=\Xw\approx2.1\times10^7$ respectively. These 
properties are summarized in \reftab{prop}, where we have also introduced 
$\Xs\approx2.3\times10^7$ as the value of $X$ for which the sound speed 
vanishes. From Eqs.~(\ref{wk}) and (\ref{csk}), this also corresponds to the 
value where the equation of state parameter $\wk=-1$. 
\begin{table}[t] 
\begin{tabular}{|l|c|c|c|c|c|c|c|} 
\hline 
\hline 
    & $(0,\Xc)$ & $\Xc$ & $(\Xc,\Xw)$ & $\Xw$ & $(\Xw,\Xs)$ & $\Xs$ & $\Xs<X$ \\ 
\hline 
$\ek$  & $+$ & $+$   & $+$ & $0$   & $-$ & $-$ & $-$ \\ 
$\wk+1$ & $+$ & $+$   & $+$ & $\infty$ & $-$ & $0$ & $+$ \\ 
$\csk$ & $+$ & $\infty$ & $-$ & $-$   & $-$ & $0$ & $+$ \\ 
\hline 
\hline 
\end{tabular} 
\caption{This table shows the sign of some variables as a function of $X$ for 
the two models analyzed in section~\ref{analysis}. For each model the values of 
$\Xc$, $\Xw$ and $\Xs$ are given in the text. The symbol ``$\infty$'' stands for 
a diverging value.} 
\label{prop} 
\end{table} 
In what follows we only analyze the dynamics for $X<\Xc$, which 
corresponds to $y>\yc\approx2.4\times10^{-4}$, since in this case $\ek>0$, 
$\wk>-1$ and $\csk>0$ and also the solution which ``solves'' the coincidence 
problem is in this region of the phase diagram. 
 
\begin{figure}[t] 
\includegraphics[scale=0.38,angle=-90]{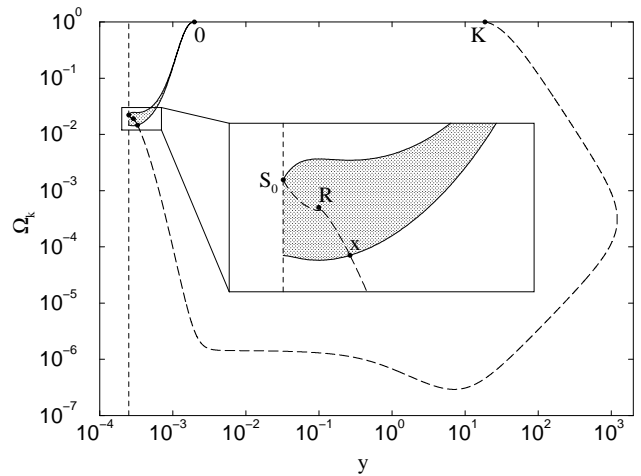} 
\caption{Basin of attraction (shaded region) of the tracker solution $\R$ during 
radiation domination for the first model analyzed in Section~\ref{analysis}. The 
points $\K$ and $\xx$ denote the k-essence domination attractor and the saddle 
point respectively, defined in Ref.~\cite{AMS2}. The solid lines are the 
limiting solutions which demarcate the basin of attraction and the long-dashed 
line separates the trajectories originating from $\OO$ and the ones originating 
from the singularity. The vertical dashed line is the singularity $y=\yc$. $\So$ 
denotes the point $(\yc,\Okc)$.}\label{basin_1} 
\end{figure} 
  
Since our concern over the dependency on initial conditions really only relates 
to the period of radiation domination, we assume that the Universe is filled 
with radiation ($\wf=1/3$). We have run simulations in order to determine the 
size of the basin of attraction of the tracker $\R$. In \reffig{basin_1} we have 
plotted the phase diagram for the k-essence field during radiation domination. 
The solid and long-dashed lines are the limiting solutions which separate the 
different types of trajectories. The vertical dashed line is the singularity 
$y=\yc$. As explained in section~\ref{formalism}, it can be divided in two 
parts: $\Sp$ (below $\So$) and $\Sm$ (above $\So$). All the trajectories 
contained within the long-dashed line originate from the same point $\OO$ and 
the rest originate from segment $\Sp$. The solid lines separate the trajectories 
ending on $\Sm$, the ones reaching the tracker solution $\R$ and the ones 
reaching the k-essence domination attractor $\K$. (Having a look at the 
trajectories plotted in \reffig{basin_2} can be useful for understanding the 
dynamics, although this figure describes the model we study next.) As we can 
see, the basin of attraction of $\R$ (shaded region) is very small and most of 
the solutions reach $\K$. However as this is still during radiation domination, 
they reach the k-essence domination attractor too early (typically after about 
$12$ e-foldings, i.e. after an increase of the scale factor $a(t)$ by about a 
factor $10^5$) to be associated with the onset of matter--radiation equality and 
to be a candidate for dark energy. We also note that in a small region of the 
phase diagram the solutions cease to exist after a finite time as they reach 
$\Sm$, the part of singularity $y=\yc$ above $\So$.

Therefore, when analyzed in detail, this first model exhibits a rather different
phase diagram structure from that sketched in Fig.~3 of Ref.~\cite{AMS2} to
generically describe models of class $(\mathrm{B}_r)$.  Unfortunately a complete 
comparison
is not possible as the precise equations used to generate that figure are not
disclosed, so that we do not know to which k-essence model the figure 
corresponds (nor whether the model has an explicit lagrangian description or is 
only of a more phenomenological nature).

\subsection{Model 2 (Ref.~[8])}

This second model is defined by 
\begin{equation} 
\pti(X) \equiv -2.05+2\sqrt{1+f(X)} \comma 
\end{equation} 
where 
\begin{equation}
\label{model2}
\begin{split} 
f(X) \equiv& \,X-10^{-8}X^2+10^{-12}X^3-10^{-16}X^4 \\ 
      & +10^{-20}X^5-10^{-24}X^6/2^6 \period 
\end{split} 
\end{equation} 
As in the previous example, this model is not always well defined. 
For $X>X_\mathrm{max}\simeq6.3\times10^{5}$ we have $1+f(X)<0$ and 
therefore $\pti(X)$ becomes ill-defined. Moreover, for $X<X_\mathrm{max}$, this 
model has the same problems as those of the model we have analyzed above. As 
before, its properties are summarized in \reftab{prop}, where this time 
$\Xc\approx3.9\times10^{5}$, $\Xw\approx5.0\times10^{5}$ and 
$\Xs\approx5.3\times10^{5}$.
 
\begin{figure}[t] 
\includegraphics[scale=0.38,angle=-90]{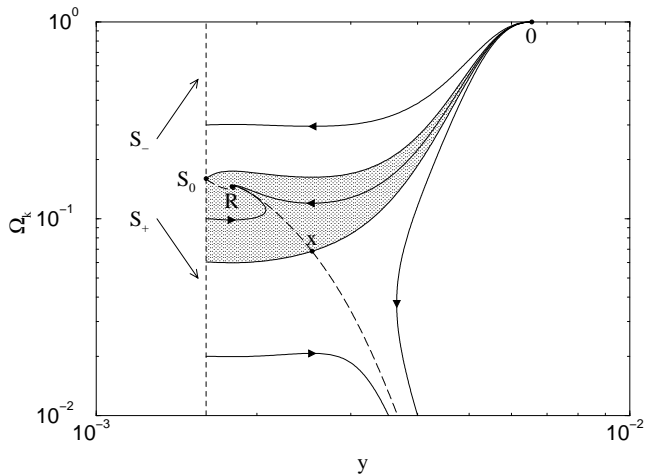} 
\caption{Basin of attraction (shaded region) of the tracker solution $\R$ during 
radiation domination for the second model analyzed in Section~\ref{analysis}. 
The explanations are the same as for \reffig{basin_1}. $\Sp$ and $\Sm$ 
denote the parts of the singularity from where some trajectories originate and 
end respectively. A few trajectories have been plotted.} 
\label{basin_2} 
\end{figure} 

As in the first example, this is a model of type $(\mathrm{B}_r)$ and for 
$y>\yc\approx1.6\times10^{-3}$ its phase space has a similar structure to that 
shown in \reffig{basin_1}. In \reffig{basin_2} we have plotted the basin of 
attraction of the tracker solution $\R$ during radiation domination for this 
second model. In order to help in understanding the dynamics a few trajectories 
have been added. Again, we see that for most of the initial conditions the field 
does not reach the tracker, but instead it reaches either an early k-essence 
domination solution $\K$ or the singularity on $\Sm$. Therefore, this second 
model suffers the same fine-tuning as that of the first example.

\section{Discussion}

As introduced in Refs.~\cite{AMS1,AMS2}, the idea of k-essence seems appealing. 
Current models of quintessence suffer in general because of the 
fine-tuning of the potential parameters to account for the fact that the field 
has only recently started dominating. K-essence was introduced as an extension 
of quintessence models by taking into account non-canonical kinetic terms. In a 
subclass of these models~\cite{AMS1,AMS2} a tracker behaviour occurs during 
radiation domination, and a cosmological-constant-like behaviour shortly after 
the transition to matter domination. Since the k-essence field seems to change 
its behaviour generically 
for purely dynamical reasons, these models could be claimed to solve the 
coincidence problem without fine-tuning of the initial conditions. In that sense 
they are more natural 
than quintessence models which rely on a different, non-obvious, scale for the 
transition to occur.

In this short paper, we have addressed in a bit more detail the question over 
the nature of this attractor solution during radiation domination which is meant 
to avoid a fine-tuning of the initial conditions. We have numerically solved the 
evolution equations for the k-essence fields and written the results in terms of 
the physically motivated parameters $y$ and $\Ok$, following 
Refs.~\cite{AMS1,AMS2}. The key result we have found is that the basin of 
attraction for the tracker solution appears to be very small compared to the 
basin of attraction for the k-essence domination solution and therefore it 
cannot be seen as equivalent to that of quintessence models. In other words, for 
almost all initial conditions, the system would evolve rapidly into k-essence 
domination. It would have done so way before matter--radiation equality and in 
that sense, the required behaviour of k-essence can only hold for a specific 
subset of initial conditions. If these turn out to be an important set of 
conditions then k-essence can be thought of as providing an elegant way of 
obtaining the acceleration we see today. However, if there is no particular 
reason for choosing such initial conditions, then we believe that k-essence 
suffers from the same fine-tuning issues that plague quintessence models.

With regard to this issue, in Refs.~\cite{AMS1,AMS2} the authors argue that the 
basin of attraction of the radiation attractor of the two models studied in this 
paper are compatible with equipartition. This particular initial condition may 
be the case should the k-essence or quintessence field be associated with one of 
the many fields produced at the end of a period of inflation. However, it need 
not be the case, and if it was not, equipartition would not help in choosing 
initial conditions. Indeed, equipartition is generally applied to systems where 
particle production occurs after the decay of the inflaton, whereas in the case 
of vacuum energy the initial condition may well have been set by early Universe 
physics (see for example Ref.~\cite{ML}). In any case, both basins of attraction 
are so small that their main parts do not overlap and therefore there remains a 
fine-tuning issue even when assuming equipartition.

We have also shown that the models proposed so far feature a singularity 
associated with a diverging sound speed. This leads to problems like the sudden 
disappearance of some field trajectories. The presence of this singularity could 
be avoided by adding an extra term to the functions given by Eqs.~(\ref{model1}) 
and (\ref{model2}), but our attempts to do so have led to the creation of a 
second radiation attractor which goes against the goal of having a model for 
which the late-time behaviour is independent of the initial conditions. Another 
way of solving this problem would be to argue that the theory is valid up to a 
certain cut-off which excludes the singularity, but as long as the radiation 
tracker is very close to the singularity this argument cannot be applied.

In this paper we have dealt with the specific lagrangians given in 
Refs.~\cite{AMS1,AMS2} at face value. Those lagrangians have a complicated form 
which is not appealing, and it is disappointing that so far it has not proved 
possible to produce much simpler models, as one would expect to be able to if 
the desired triggering behaviour really is generic to k-essence. The particular 
nature of the models makes it hard
to assess how difficult it is to build a working example, for instance 
quite how much fine-tuning is required to establish the time needed between 
matter--radiation equality and k-essence domination, and how easily one can 
avoid the presence of many different attractors during radiation domination.
Unfortunately, we are not aware of any 
particular particle physics motivated models which would deliver the lagrangians 
of the two models, and, as with quintessence, we believe that a realistic model 
of 
k-essence remains a challenge which has to be met.


\begin{acknowledgments} 
M.M.~was supported by the Fondation Barbour, the Fondation Wilsdorf, the 
Janggen-P\"{o}hn-Stiftung and the ORS Awards Scheme, and A.R.L.~in part by the 
Leverhulme Trust. We thank Mark Trodden for many discussions relating to 
k-essence. 
\end{acknowledgments}


\end{document}